\documentclass[a4paper, amsfonts, amssymb, amsmath, reprint, showkeys, nofootinbib, twoside]{revtex4-1}
\PassOptionsToPackage{colorlinks=true,linkcolor=blue,citecolor=blue}{hyperref}
\usepackage{tikz}
\usepackage{tikz-feynman}
\usepackage[english]{babel}
\usepackage[utf8]{inputenc}
\usepackage{orcidlink}
\usepackage{lipsum}
\usepackage[colorinlistoftodos, color=green!40, prependcaption]{todonotes}
\usepackage{tabularx}

\usepackage{braket}

\usepackage{hyperref}
\begin{document}

\title{$U(1)_{B-L}$ Dark Matter Constrains Smooth (SUSY) Hybrid Inflation}

\author{Karim M. Selim,\orcidlink{0009-0001-5601-4562}}
\email{karimselim@gstd.sci.cu.edu.eg}
\email{karimselim322@gmail.com}
\affiliation{Cairo University, Department of Physics, Giza, Cairo, Egypt}

\author{A.Y Ellithi}
\email{aliellithi@cu.edu.eg}
\affiliation{Cairo University, Department of Physics, Giza, Cairo, Egypt}

\author{M. Abolmahassen}
\affiliation{Cairo University, Department of Physics, Giza, Cairo, Egypt}

\date{\today}

\begin{abstract}
We propose a unified framework that connects inflationary dynamics with dark matter production within a supersymmetric $U(1)_{B-L}$ extension of the Standard Model. The setup is based on smooth hybrid inflation embedded in supergravity, with a non-minimal K\"ahler potential ensuring the control of higher-order corrections.

The model involves three scalar fields: the inflaton $\sigma$, an auxiliary field $\zeta$ responsible for ending inflation, and a singlet mediator $\eta$ that links the inflationary and dark sectors. Dark matter is realized as an inert scalar stabilized by a $\mathbb{Z}_2$ symmetry and produced non-thermally via reheating dynamics.
The model is also constructed within the framework of supergravity, incorporating corrections from a non-minimal K\"ahler potential~\cite{ref1}. This extension is essential to achieve inflationary observables consistent with current data while keeping higher-order corrections under control. In the minimal scenario, the spectral index tends to approach unity ($n_s \to 1$), which is in tension with observational constraints, particularly those related to primordial gravitational waves. Therefore, the inclusion of non-minimal K\"ahler corrections becomes necessary to obtain a viable inflationary model.

We show that the requirement of reproducing the observed dark matter relic abundance imposes strong constraints on the inflationary sector, significantly reducing the allowed parameter space. As a result, the scalar spectral index is tightly constrained to $n_s \simeq 0.972 - 0.974$, consistent with current observational bounds.

While the inflaton–dark matter coupling has a negligible effect on the background evolution, it induces observable modifications in the tensor-to-scalar ratio and the spectrum of primordial gravitational waves. This establishes a direct link between dark matter physics and inflationary observables.
\end{abstract}


\maketitle

% تأكد إن كل الملفات دي موجودة بنفس الاسم
\section{Model Setup}

We consider a supersymmetric framework including supergravity corrections, in which the dynamics of inflation are directly connected to non-thermal dark matter production. The model contains three scalar fields: the inflaton $\sigma$, an auxiliary field $\zeta$ responsible for terminating inflation, and a singlet mediator $\eta$ that governs the transfer of energy to the dark sector.

Although the theory is intrinsically multi-field, the inflationary dynamics effectively reduce to a single-field trajectory. This occurs due to the stabilization of the directions orthogonal to the inflationary path in field space. After the end of inflation, the inflaton oscillates around the minimum of the potential and transfers its energy to the mediator field $\eta$, which subsequently produces dark matter particles through $\eta$-mediated interactions. This provides a consistent realization of non-thermal dark matter production tied to the reheating dynamics.

\subsection{Scalar Potential}

The scalar potential governing the inflationary sector is given by [1]
\begin{equation}
 V = \left[\frac{\zeta^4}{16M_S^2}-M_X^2\right]^2 + \frac{\sigma^2 \zeta^6}{16M_S^4},
\end{equation}
where $\sigma$ is the inflaton field and $\zeta$ is the auxiliary field. The supersymmetric vacuum is located at
\begin{equation}
\sigma = 0, \quad \frac{\zeta}{2} = \sqrt{M_X M_S}.
\end{equation}

Minimizing the potential along the inflationary trajectory yields
\begin{equation}
\zeta = \sqrt{6}\sigma \left[\sqrt{1+\frac{\chi^4}{36\sigma^4}} -1\right]^{1/2},
\end{equation}
with $\chi = 2\sqrt{M_X M_S}$.

The full scalar potential, including interactions with the mediator $\eta$, 

\begin{equation}
\begin{aligned}
V=\;& -\mu_1^2|\sigma|^2+\mu_2^2|\zeta^2|-\mu_\eta^2 |\eta|^2+\lambda_1 |\sigma|^4 + \lambda_2 |\zeta|^4 + \lambda_3|\eta|^2 |\zeta|^{2}+  \\&\lambda_4  |\sigma^\dagger \zeta|^2 + \lambda_\eta |\eta|^{4} +  \kappa_1|\eta|^2 |\sigma|^{2} +  \kappa_2 |\sigma|^2 |\zeta|^2 \\
&+ \frac{\lambda_5}{2}\left(\sigma^{\dagger}\zeta\, \sigma^{\dagger}\zeta + \zeta^{\dagger}\sigma\, \zeta^{\dagger}\sigma \right).
\end{aligned}
\end{equation}

\subsection{Yukawa Sector}

The Yukawa Lagrangian is given by
\begin{align}
\mathcal{L}_{\text{Yukawa}} =\;& -Y_u^{ij}\bar{Q}^{i}_{L}\tilde{\sigma} u_R^{j}
- Y_d^{ij}\bar{Q}^{i}_{L}\sigma d_R^{j} \nonumber \\
& - Y_e^{ij}\bar{L}^{i}_{L}\sigma e_R^{j}
- Y_\nu^{ij}\bar{L}^{i}_{L}\tilde{\sigma} \nu_R^{j} \nonumber \\
& - Y_\eta^{ij}\bar{\nu}^{i}_{R}\eta (\nu_R^{j})^{c}
+ \text{h.c.}
\end{align}

\subsection{Effective Masses}

The effective mass of the auxiliary field $\zeta$ is obtained from the second derivative of the scalar potential prior to imposing the inflationary trajectory, Fig[1]
\begin{equation}
M_\zeta^2 = \frac{\partial^2 V}{\partial \zeta^2},
\end{equation}
which yields

\begin{equation}
\begin{split}
M_\zeta^2 =\;& 2 \left[\frac{\sqrt{6}\, \sigma^{3} \left(6 \sqrt{36+\frac{\chi^{4}}{\sigma^{4}}}-36\right)^{\frac{3}{2}}}{144 M_S^{2}}\right]^{2} \\
&+ 2 \left[\frac{\sigma^{4} \left(6 \sqrt{36+\frac{\chi^{4}}{\sigma^{4}}}-36\right)^{2}}{576 M_S^{2}} - M_X^{2}\right] \\
&\times \left[\frac{\sigma^{2} \left(6 \sqrt{36+\frac{\chi^{4}}{\sigma^{4}}}-36\right)}{8 M_S^{2}}\right] \\
&+ \frac{5 \sigma^{6} \left(6 \sqrt{36+\frac{\chi^{4}}{\sigma^{4}}}-36\right)^{2}}{96 M_S^{4}} \, .
\end{split}
\end{equation}

\begin{figure}
    \centering
    \includegraphics[width=1\linewidth]{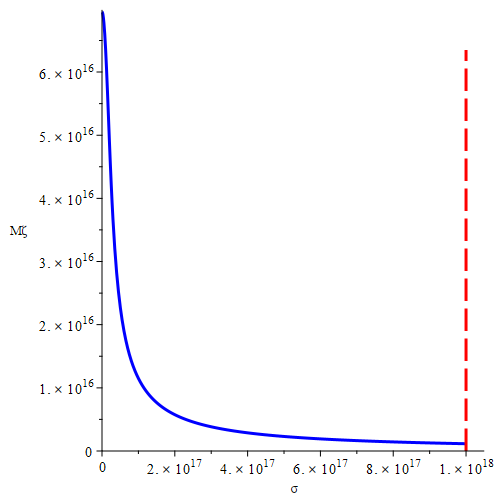}
   \caption{Auxiliary field mass vs inflation field  ensuring it within the cutoff scale (All units in GeV)}
    \label{fig:placeholder}
\end{figure} 
Similarly, the inflaton mass is
\begin{equation}
M_\sigma^2 = \frac{\sigma^{6} \left(6 \sqrt{36+\frac{\chi^{4}}{\sigma^{4}}}-36\right)^{3}}{1728 M_S^{4}}.
\end{equation}

\subsection{Inflaton Potential with Supergravity Corrections}

Including non-minimal K\"ahler corrections
\begin{equation}
    K = |S|^2 +|\Delta_R|^2+ |\Bar{\Delta}_R|^2 + \frac{\kappa_s |S|^4}{4 M_p^2}
\end{equation}
with $ \Delta_R = \Bar{\Delta}_R$ and $\sigma^2 >> M^2$
, the inflaton potential becomes [1]

\begin{equation}
V = M_X ^4 \left[1-\kappa_s \frac{\sigma^2}{2M_p^2}
+(1-\frac{7}{2}\kappa_s+2\kappa_s^2)\frac{\sigma^4}{8M_p^4}
-\frac{2}{27} \frac{M^4}{\sigma^4}\right].
\end{equation}
\subsection{Benchmark Parameters}

We summarize representative benchmark values for the inflationary parameters in Table~\ref{tab:inflation_params}.
for slow-rolling parameters ,
$\epsilon = \frac{M_{Pl}^2}{16\pi}(\frac{V'}{V})^2, \eta = \frac{M_{Pl}^2}{8 \pi}\frac{V''}{V}$ and $|\eta| =1$
\begin{table}[h]
\centering
\begin{tabular}{|c| c|c|c|c|c|}
\hline\hline
Set & $\kappa_s$ & $M$ (GeV) & $M_X$ (GeV) & $\sigma_{\rm end}$ (GeV) \\
\hline
1 & $10^{-5}$ & $7.3\times10^{17}$ & $7.3\times10^{17}$ & $1.1\times10^{18}$ \\
2 & $0.02$   & $7.3\times10^{17}$ & $7.3\times10^{17}$ & $1.14\times10^{18}$ \\
\hline\hline
\end{tabular}
\caption{Benchmark parameter sets used in the analysis of the inflationary dynamics.}
\label{tab:inflation_params}
\end{table}
\subsection{Dark Matter Sector}

Expanding around the vacuum expectation values (VEVs), 
\begin{equation}
\langle \sigma \rangle = \frac{1}{\sqrt{2}}
\begin{pmatrix}
0 \\ v
\end{pmatrix}, \quad
\langle \zeta \rangle = 0, \quad
\langle \eta \rangle = v_\eta,
\end{equation}
These VEVs does not occur simultaneously with Eq[2]  since that the Dark matter production is taken a place after the end of inflation
the dark matter mass is given by
\begin{equation}
m^2_{\zeta} =\frac{\kappa_2 v_\eta^2}{2} + \lambda_l v^2 + \mu_2^{2},
\end{equation}
where $\lambda_l = \frac{1}{2}(\lambda_3+\lambda_4-\lambda_5)$.

For representative parameter values,
\begin{equation}
\lambda_l =0.129, \quad \lambda_1=0.01, \quad \lambda_3 =1.003,
\end{equation}
and $\kappa_2 \in [10^{-5}, 0.02]$, we obtain dark matter masses in the range
\begin{equation}
m_\chi \sim 40 - 1000~\text{GeV}.
\end{equation}

\subsection{Relic Abundance}

The relic abundance is computed by solving the Boltzmann equations, [2],[3]
\begin{equation}
    \bar{H} \frac{d\Phi}{dA}=-c_\rho^{1/2} A^{1/2} \Phi
    \end{equation}
    \begin{equation}
      \bar{H}   \frac{dR}{dA}=
      c_\rho^{1/2}A^{3/2}(1- \frac{b E_\chi}{m_\sigma})+ c_1^{1/2}A^{-3/2} \sigma v E_\chi \bar{H} \frac{d\Phi}{dA}M_{Pl}( X^2-X_{eq}^2)
    \end{equation}
    \begin{equation}
        \bar{H} \frac{dX}{dA}=A^{1/2}T_{RH} \Phi \frac{b}{m_\sigma}-c_1^{1/2} A^{-5/2} \sigma vM_{Pl}T_{RH} ( X^2-X_{eq}^2)
    \end{equation}
    where $\Phi = \rho_\sigma \frac{a^3}{T_{RH}},R = \rho_r a^4, X=n_\chi a^3, A=a T_{RH}$
    and  $c_\rho =\frac{\pi^2 g\star T_{RH}}{30}, c_1 = \frac{3}{8\pi }$ and X is the scaled number density at equilibrium 
    $ X_{eq}  = (\frac{A}{T_{RH}})^3g_X (\frac{M_XT}{2\pi})^{\frac{3}{2}} e^{\frac{-M_X}{T}}$ 
    Solving Boltzmann 1st equation we get 
    \begin{equation}
        \Phi = K e^{-\frac{2\sqrt{c_\rho }A^{3/2}}{3H}}
    \end{equation}\vspace{2 cm}
    with  initial conditions $
    \Phi_I = \frac{3M_p^2H_I^2}{T^4_{RH}},R_I = X_I =0, A_I=1 $ we get
    \begin{equation}
        \Phi =\frac{3M_p^2H_I^2}{T^4_{RH}}e^{-\frac{2\sqrt{c_\rho }[A^{3/2}-1]}{3H}}
    \end{equation}
    the second and third equation we assume that $ X <<< X_{eq} $ 
    we get
   
    and \begin{equation}
        X = \frac{2}{3 \Bar{H}}T_{RH} \Phi\frac{b\braket{E_\chi }}{m_\phi} [A^{3/2}-1] -\frac{2}{3} \sqrt{c_\rho}\braket{\sigma v} M_p T_{RH } X_{eq}^2 [A^{-3/2}-1]
    \end{equation}
    \begin{equation}
R = \frac{2}{5 \bar{H}} \sqrt{c_\rho} \left[1 - \frac{b \langle E_\chi \rangle}{m_\sigma} \right] \Phi \left(A^{5/2} - 1\right) 
+ \frac{4}{\bar{H}} \sqrt{c_\rho} \, \langle \sigma v \rangle \langle E_\chi \rangle M_p X_{eq}^2 \left(A^{-1/2} - 1\right)
\end{equation}

    \begin{equation}
        \rho_r =\frac{\pi^2}{30}g_\star T^4
    \end{equation}
  
  from GR basic relation we get
  \begin{equation}
      \rho_r = \rho_{r_0}(\frac{a_0}{a})^4
  \end{equation}
  \begin{equation}
       \rho_{DM} = \rho_{DM_0}(\frac{a_0}{a})^3
  \end{equation}
   , dividing  both 24 and 25  and $\frac{a_0}{a} = \frac{T}{T_0}$
yielding
\begin{equation}
    \Omega_{DM} h^2 = \frac{\rho_X}{\rho_r}\frac{T_f}{T_{now}}\Omega_{R} h^2
\end{equation}
using the co-moving relation we get
\begin{equation}
\Omega_{DM} h^2 = M_{\chi}\frac{X_f}{R_f} A_f\frac{T_{f}}{T_{now}T_{RH}} \Omega_{R}h^2.
\end{equation}
\vspace{3 cm}
Using Eq.~(26), we perform a parameter scan in the $(T_{\rm RH},M_{\rm DM})$ plane by varying both the reheating temperature and the dark matter mass. For each point in the scan, the relic abundance is computed and compared with the observed value, $\Omega_{\rm DM} h^2 = 0.12 \pm 0.001$. The points satisfying the observed relic abundance constraint are then projected onto the $(T_{\rm RH},M_{\rm DM})$ plane, yielding the distribution shown in Fig.~9.

Figure~9 demonstrates that the model successfully reproduces the measured dark matter relic abundance within a well-defined region of the parameter space. The scan further reveals that the allowed dark matter mass range is correlated with the reheating temperature, leading to viable solutions for reheating temperatures in the approximate interval
\[
T_{\rm RH}\sim 0.6-1~{\rm GeV}.
\]
Therefore, the requirement of obtaining the correct relic abundance not only constrains the dark matter mass but also imposes nontrivial bounds on the reheating dynamics. This result establishes a direct connection between the reheating history of the Universe and the viable dark matter parameter space in the model.

 for the initial conditions 
\begin{equation}
    \Phi_I=\frac{3M_{P}^2 H_I^2}{T_{RH}^4} , R_I=X_I=0, A_I =1
\end{equation}
we obtain 
\begin{equation}
    H = \sqrt{\frac{5\pi^2g_\star^2(T)}{72g_\star (T_{RH})}}\frac{T^4}{T_{RH}^2 M_P}
\end{equation}
with the temperature 
\begin{equation}
    T =(\frac{30}{\pi^2g_\star(T)}) ^{1/4}\frac{R^\frac{1}{4}}{A} T_{RH}
\end{equation}

Fig [2] indicates to the how Reheating temperature and the density evolve with the universe expansion, Figures 3–5 are derived from the solution of the Boltzmann equation and demonstrate the evolution of comoving densities as a function of the scale factor

\begin{figure}
    \centering
    \includegraphics[width=1\linewidth]{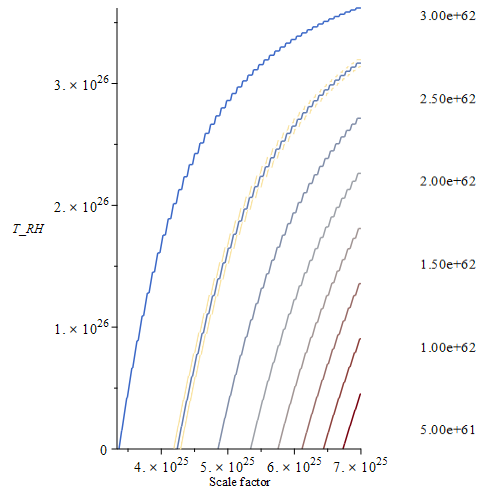}
    \caption{Reheating Temperature vs Scale Factor Corresponds to Co-moving Density}
    \label{fig:placeholder}
\end{figure}
\begin{figure}
    \centering
    \includegraphics[width=1\linewidth]{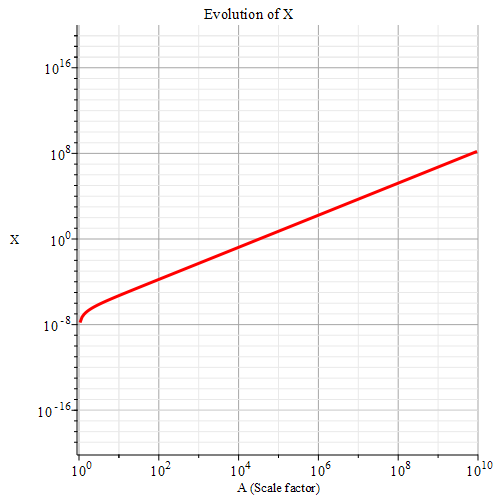}
   \caption{The DM Density Vs The Scale Factor}
    \label{fig:placeholder}
\end{figure}
\begin{figure}
    \centering
    \includegraphics[width=1\linewidth]{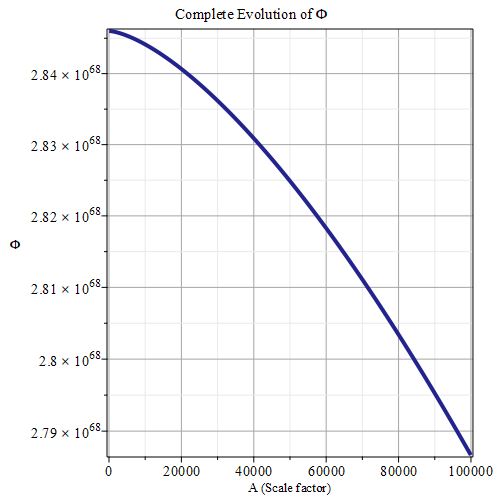}
    \caption{The $\Phi$ Vs The Scale Factor}
    \label{fig:placeholder}
\end{figure}
\begin{figure}
    \centering
    \includegraphics[width=1\linewidth]{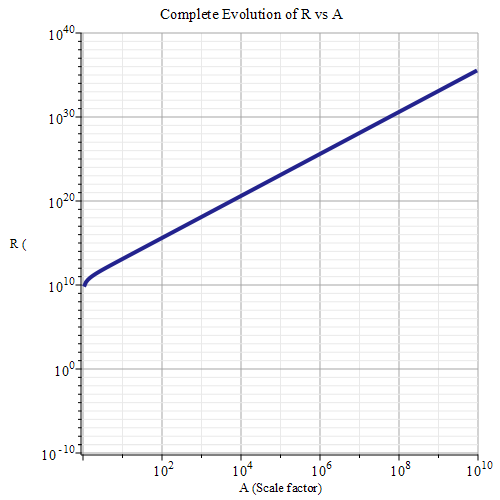}
    \caption{R vs scale factor}
    \label{fig:placeholder}
\end{figure}

\begin{figure}[h]
\centering
\begin{tikzpicture}
\begin{feynman}
\diagram {
i1 [particle=$\sigma$] -- [scalar] a -- [scalar] f1 [particle=$\eta$],
i2 [particle=$\sigma$] -- [scalar] a -- [scalar] f2 [particle=$\eta$],
};
\end{feynman}
\end{tikzpicture}
\caption{Tree-level scattering $\sigma \sigma \rightarrow \eta \eta$.}
\end{figure}
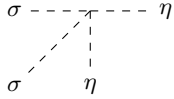
\begin{figure}[h]
\centering
\begin{tikzpicture}
\begin{feynman}
\diagram {
i1 [particle=$\eta$] -- [scalar] a -- [scalar] f1 [particle=$\zeta$],
i2 [particle=$\eta$] -- [scalar] a -- [scalar] f2 [particle=$\zeta$],
};
\end{feynman}
\end{tikzpicture}
\caption{Tree-level annihilation $\eta \eta \rightarrow \zeta \zeta$.}
\end{figure}
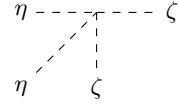
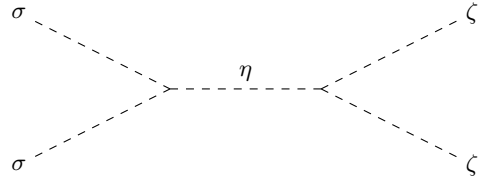
\begin{figure}[h]
\centering
\begin{tikzpicture}
\begin{feynman}
\vertex (a) at (-2,1) {$\sigma$};
\vertex (b) at (-2,-1) {$\sigma$};
\vertex (c) at (0,0);
\vertex (d) at (2,0);
\vertex (e) at (4,1) {$\zeta$};
\vertex (f) at (4,-1) {$\zeta$};

\diagram*{
(a) -- [scalar] (c),
(b) -- [scalar] (c),
(c) -- [scalar, edge label=$\eta$] (d),
(d) -- [scalar] (e),
(d) -- [scalar] (f),
};
\end{feynman}
\end{tikzpicture}

\caption{$\sigma\sigma \to \eta \eta \to \zeta\zeta$}
\end{figure}
We identify reheating with the process
\begin{equation}
\sigma \sigma \rightarrow \eta\eta \rightarrow \zeta \zeta
\end{equation}

The production of the dark matter field $\zeta$ from the inflaton $\sigma$ proceeds via a two-step scattering process, $\sigma \sigma \rightarrow \eta \eta \rightarrow \zeta \zeta$, where $\eta$ acts as an intermediate mediator.
. 
\begin{equation}
    \sigma v = \frac{1}{2M_\sigma ^2} \int\frac{d^3p}{(2\pi)^3}\int \frac{d^3p'}  {(2\pi)^3} \frac{1}{4M_\zeta^2} |\mathcal{M}|^2 (2\pi)^4 \delta^4(M_\sigma-(p+p'))
\end{equation}
\vspace{2 cm}
The scattering amplitude for each diagrams can be written as follows :
\begin{equation}
   - i\mathcal{M}_{\sigma \sigma\to \zeta \zeta } = -i\kappa_2
\end{equation}
\begin{equation}
   - i\mathcal{M}_{\eta \eta \to \zeta \zeta \ } = -i \lambda_3
\end{equation}
\begin{equation}
   - i\mathcal{M}_{\sigma \sigma\to \eta \eta  \to \zeta \zeta  } = -i\frac{\kappa_1 \lambda_3}{P_\eta^2-M_{\eta}^2}
\end{equation}
b integrating assuming all field are non-relativistic i.e$ p \approx M$
leading to
\begin{equation}
    \sigma v_{\sigma\sigma\to \zeta\zeta}\approx
\frac{1 }{64\pi M_\sigma^2}\kappa_2^2 = 1.8\times10^{-44} GeV^{-2}
\end{equation}
\begin{equation}
    \sigma v_{\sigma\sigma\to \eta\eta}\approx
\frac{1 }{64\pi  M_\sigma^2}\kappa_1^2=2.94\times10^{-36} GeV^{-2}
\end{equation}
\begin{equation}
\sigma v_{\eta\eta\to \zeta\zeta}\approx
\frac{1 }{64\pi M_\eta^2}\lambda_3^2 = 5 \times10^{-11} GeV^{-2}
\end{equation}
where at $M_X = M_S =1.7\times10^{17}$GeV $M_\sigma=5.3\times10^{15}$ GeV, $M_\zeta = 7 \times10^{16}$ GeV
 From the scalar potential, two interaction channels involving the inflaton field $\zeta$ can be identified. 
The first corresponds to a direct interaction channel responsible for the dark matter imprint on cosmological observables, 
while the second proceeds via an intermediate two-step interaction mediated by the $\eta$ field. 
Both contributions are included in the evaluation of the scattering amplitude and the corresponding cross section. 
Nevertheless,the direct channel is strongly suppressed and gives a negligible contribution due to the large hierarchy 
$M_\sigma \gg M_\eta$ entering the  denominator. Although the direct interaction is subdominant in the production of dark matter, the associated coupling $\kappa_2$ is present at the Lagrangian level and therefore enters the fundamental dynamics of the model. In particular, it can contribute to the effective inflaton potential and thus affect the slow-roll parameters, leading to an indirect dependence of the spectral index on $\kappa_2$.

At later stages, the same coupling governs the direct production channel, which remains suppressed compared to the mediator-induced process. This highlights the distinct roles played by the same interaction across different cosmological epochs.
$\sigma v\approx 5\times 10^{-11}$ $(GeV)^{-2}$

\subsection{EFT Validity}

Requiring the validity of the effective field theory throughout inflation imposes constraints on the parameter space. In particular, demanding that the cutoff scale does not exceed the Planck scale leads to
\begin{equation}
\sigma \lesssim M_p,
\end{equation}
which is satisfied in the viable regions of parameter space.
\begin{figure}
    \centering
    \includegraphics[width=1\linewidth]{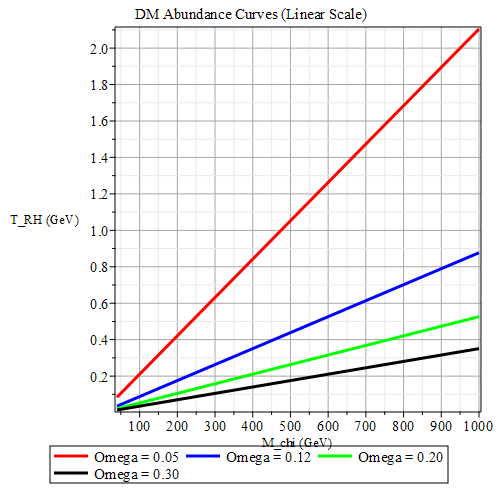}
    \caption{ blind scan for which parameters can achieve the observable Relic within $ \Omega h^2$ Vs DM mass with Reheating Temperature All Units in GeV with $M_\sigma = 1.7\times 10^{10} GeV , b= 1\times10^{-9}$}
    \label{fig:placeholder}
\end{figure}
It is worth mentioning that the field mass scales inversely with the field value, ensuring that the model remains within the EFT regime at sufficiently high masses, as shown in Fig.~1.

To assess the validity of the effective field theory (EFT) during inflation, we examine the evolution of the auxiliary field mass $M_\zeta$ along the inflationary trajectory. In Fig.~1, we plot $M_\zeta$ as a function of the inflaton field $\sigma$, together with the EFT cutoff scale, taken to be the reduced Planck mass $M_p$.

The inflationary regime is identified by the interval $\sigma_Q \rightarrow \sigma_{\text{end}}$, where $\sigma_Q$ corresponds to the field value at horizon crossing and $\sigma_{\text{end}}$ marks the end of inflation. These values are extracted from the benchmark parameter sets listed in Table~II and are explicitly indicated in the figure.

Within this interval, we observe that the auxiliary field mass exhibits a nontrivial evolution and may approach or exceed the cutoff scale $M_p$. The region where $M_\zeta \gtrsim M_p$ signals a breakdown of the EFT description, as higher-dimensional operators can no longer be neglected. This behavior is associated with a localized peak in the mass profile, reflecting a rapid increase in the curvature of the scalar potential.
\section{The parameter space} \label{sec:develop}

nThe mass scale $M$ is chosen to be $M = 10^{18}$ GeV, 
serving as a UV cutoff that suppresses higher-order 
corrections arising from the dark matter sector 
interactions. This ensures the validity of the EFT 
description throughout the inflationary trajectory,

compared to the leading inflationary potential.
In our analysis, the mass scales $M$ and $M_X$ are fixed consistently within the inflationary sector in order to determine the field value at horizon crossing $\sigma_Q$, as well as the corresponding predictions for the scalar spectral index $n_s$ and the tensor-to-scalar ratio $r$.

The inflationary trajectory is computed by imposing the standard number of e-folds $N \simeq 57$, while simultaneously incorporating the constraints arising from the requirement of reproducing the observed dark matter relic abundance. This procedure allows us to consistently relate inflationary observables to the dark matter sector parameters within the viable region of the model parameter space.

The choice $N \simeq 57$ corresponds to a typical reheating history compatible with the inflationary scale considered in this work.
For $\kappa_2 = 1\times 10^{-5}$ , we compute the auxiliary field for each benchmark point listed in Tables~[1] and [2]. 
it is to be noted that due to the prime EFT issue of Table I we obliged to change the parameters to be in the EFT zone that is based  implicitly on  DM Mass requirements 

We find that $M_\zeta \simeq 5.6\times10^{17}\,\mathrm{GeV}$ and $M_\zeta \simeq 7.5\times10^{16}\,\mathrm{GeV}$ for Tables~[1] and [2], respectively. 

This result highlights an intriguing feature of the model: the same field responsible for dark matter production also plays a crucial role in controlling the validity of the effective field theory (EFT). \vspace{3 CM}
This result suggests a nontrivial interplay in the model, where the field governing dark matter production simultaneously sets the scale for EFT validity.
It is important to note that the parameters $M_X$ and $M_S$ are not chosen independently, but are fixed consistently with the coupling structure in order to reproduce viable inflationary observables and maintain the validity of the effective field theory. 

As a result, the scale $M_\zeta$ is not a free parameter, but is implicitly correlated with the couplings (in particular $\kappa_2$), leading to a nontrivial interplay between dark matter production and EFT consistency.
Interestingly, the auxiliary field responsible for mediating dark matter production exhibits a nontrivial behavior along the inflationary trajectory, generating two distinct mass scales. While one of these scales remains within the regime of validity of the effective field theory (EFT), the second develops a peak that exceeds the cutoff scale, potentially signaling a breakdown of the EFT description.

This feature underscores the dual role of the auxiliary field. On the one hand, it governs reheating dynamics and facilitates non-thermal dark matter production. On the other hand, it acts as a probe of the theoretical consistency of the model, as its mass evolution directly reflects the limits of the EFT framework.

As a result, the requirement of successful dark matter production acquires a deeper significance. It does not merely impose phenomenological constraints, but effectively serves as a selection principle that restricts the parameter space to regions where the EFT remains valid throughout the inflationary evolution.

This interplay highlights that the viability of the model is highly sensitive to the choice of couplings and vacuum expectation values. In particular, a careful tuning is required to suppress the emergence of the super-cutoff mass peak, while simultaneously preserving the correct dark matter relic abundance,

In particular, the appearance of a super-cutoff mass peak reflects a rapid growth in the curvature of the scalar potential in specific regions of field space. While this does not indicate an instability of the potential itself, it signals that the corresponding regime lies beyond the domain of validity of the effective field theory.

This observation strengthens the role of the auxiliary field as a diagnostic tool for EFT consistency. Since the same field also mediates the production of dark matter, the requirement of obtaining the correct relic abundance dynamically restricts the inflationary trajectory to regions where the EFT description remains valid.

Therefore, the dark matter constraint acts as a nontrivial selection principle that eliminates regions of field space associated with excessive curvature scales, ensuring the internal consistency of the model already at the tree Level 

\begin{table}[h]
\centering
\begin{tabular}{|| c| c |c |c| c |c| c ||} 
 \hline
 $\kappa_s$ & $M$ & $M_X$ & $\sigma_{end}$ & $\sigma_Q$ & $r$ & $n_s$ \\ [0.5ex] 
 \hline\hline
 $10^{-5}$ & $1.7\times10^{17}$ & $1.7\times10^{17}$ & $4.374\times10^{17}$ & $5.52\times10^{17}$ & $1\times 10^{-5}$ & $0.972$ \\ 
 \hline
 $0.02$ & $1.7\times10^{17}$ & $1.7\times10^{17}$ & $4.4\times 10^{17}$ & $5.5\times10^{17}$ & $6\times10^{-6}$ & $0.974$ \\ [1ex] 
 \hline   
\end{tabular}

\caption{Benchmark parameter sets used in the analysis of the inflationary dynamics within EFT. All units of fields and masses are in GeV.}
\label{tab:inflation_params}

\end{table}
\section{The Results} \label{sec:outline}
\begin{figure}
    \centering
    \includegraphics[width=1\linewidth]{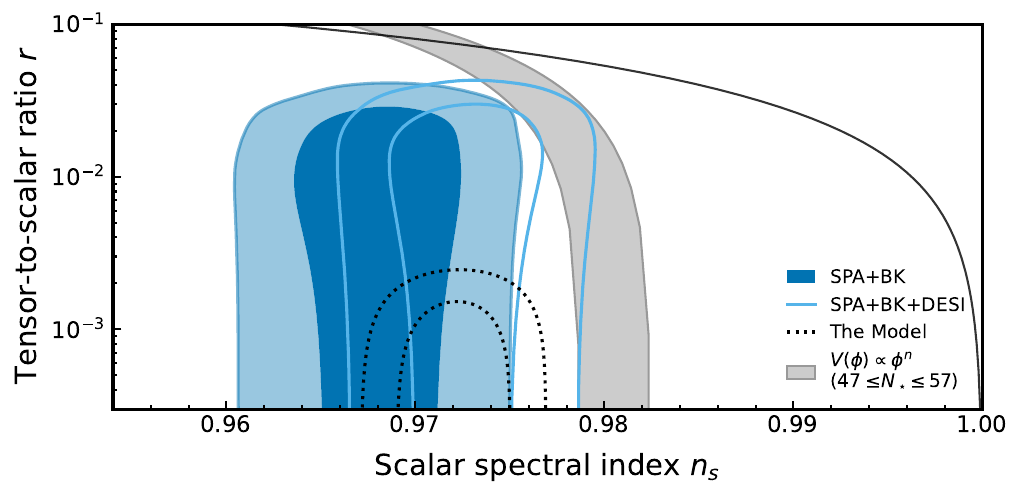}
    \caption{with coupling $\kappa_s= 10^{-5}$ }
    \label{fig:placeholder}
    \includegraphics[width=1\linewidth]{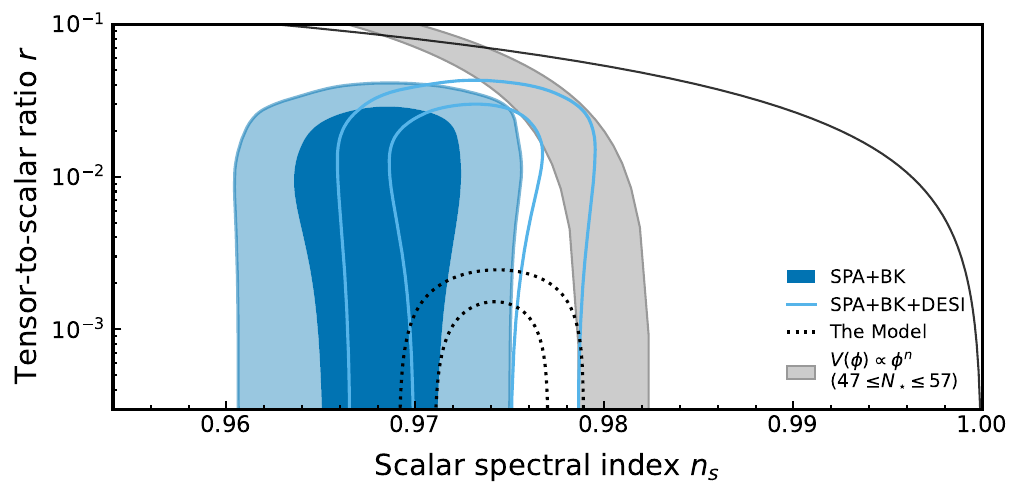}
    \caption{with coupling $\kappa_s = 0.02$ }
\end{figure}
We estimate the effective mass of the auxiliary field $\zeta$ from the curvature of the scalar potential along the inflationary trajectory. We find 
\[
m_\zeta \sim \mathcal{O}(10^{16})~\text{GeV},
\]
which is significantly larger than the Hubble scale during inflation, $H \sim \mathcal{O}(10^{14})~\text{GeV}$.  This implies
\[
\frac{m_\zeta^2}{H^2} \gg 1,
\]

ensuring that the $\zeta$ field is strongly stabilized and that the inflationary dynamics effectively that is visualized in Fig[1] reduce to a single-field description
Our analysis reveals that the requirement of reproducing the observed dark matter relic abundance leads to strong correlations among the model parameters that is visualized in Fig[2]- Fig[5]. Figures [6-8] demonstrate the mechanism of DM generations through possible ways. These constraints significantly restrict the viable parameter space and directly impact the inflationary predictions. We find that the coupling  between  the inflaton and the dark matter sector does not substantially 
modify the background evolution, in particular the field value at  the end of inflation. However, it plays a crucial role in shaping the spectrum of primordial perturbations.In particular Fig[10] and [11], we observe that small variations in the coupling parameters can induce noticeable shifts in the scalar spectral index and the tensor-to-scalar ratio. This demonstrates that dark matter physics can leave an observable imprint on inflationary cosmology.
This demonstrates that dark matter constraints act as a selection principle for inflationary models.

Evolution of the auxiliary-field mass along the inflationary trajectory. The large mass of the orthogonal field ensures its decoupling from the inflationary dynamics, thereby justifying the effective single-field description used throughout the analysis.
\section{Conclusion} \label{sec:conclusion}

We have presented a unified framework linking inflation and dark matter production within a supersymmetric $U(1)_{B-L}$ 
extension. Our analysis shows that the requirement of successful dark matter generation imposes nontrivial constraints on the inflationary sector, significantly reducing the viable parameter space.
While the background dynamics remain largely unaffected,
the primordial perturbation spectrum exhibits a strong sensitivity  to the coupling between the inflaton and the dark sector. In particular, we find that dark matter production can indirectly influence the amplitude of primordial gravitational waves.
These results highlight the importance of treating inflation and dark matter within a single coherent framework and open the possibility of testing such scenarios with future cosmological observations.

\appendix


\begin{thebibliography}{99}

\bibitem{ref1}
Khalil, Shaaban and Shafi, Qaisar and Sil, Arunansu,
Phys. Rev. D 86, 073004.

\bibitem{ref2}
Binjonaid, Maien and Elsheshtawy, Ahmed and Khalil, Shaaban,
Non-thermal Dark Matter in $U(1)_{B-L}$ Extension of Inert Doublet Model,
JCAP 03 (2025) 057.

\bibitem{ref3}
Drees, Manuel and Hajkarim, Fazlollah,
Dark matter production in an early matter dominated era,
JCAP 02 (2018) 057.

\bibitem{ref4}
Han, Chengcheng,
Higgsino dark matter in a non-standard history of the universe,
Phys. Lett. B 798 (2019) 134997.

\bibitem{ref5}
L. Balkenhol et al.,
Inflation at the End of 2025: Constraints on $r$ and $n_s$ Using the Latest CMB and BAO Data,
arXiv:2512.10613.

\end{thebibliography}
\end{document}